\documentclass[final,english]{bullsrsl}

\usepackage[latin1]{inputenc}
\usepackage[T1]{fontenc}
\usepackage{natbib} 
\usepackage{graphicx}

\begin{document}
\title{Can Neutron Star Mergers Alone Explain the R-process Enrichment of the Milky Way?}
\author[corresponding]{Dany}{Vanbeveren}
\author{Nicki}{Mennekens}
\affiliation{Astronomy \& Astrophysics Research Group, Vrije Universiteit Brussel}
\correspondance{dvbevere@vub.be}
\maketitle

\begin{abstract}
\citet{Mennekens2014} studied the effect of double compact star mergers on the Galactic chemical enrichment of r-process elements. LIGO merger detections since 2015 and new r-process element yields as function of neutron star + neutron star (NS+NS) and neutron star + black hole (NS+BH) mass requires an update of the 2014 computations. The results of the update are the scope of the present paper.
\end{abstract}

\section{Introduction}
To answer the title-question we proceed as follows. The chemistry-data of long-lived stars (G-dwarfs) gives us the observed Galactic evolution of Eu as function of Fe. With a detailed Galactic chemical evolutionary code (that includes as much as possible all effects of single star evolution as well as those of close binaries) we calculate the temporal variation of the double white dwarfs (progenitors of Type Ia supernovae responsible for a significant fraction of the Fe-enrichment), the double NS binary mergers and mixed system mergers (NS+NS and NS+BH mergers) possibly responsible for a significant r-process chemical enrichment. Linking the predicted NS+NS and NS+BH population, the merger rates caused by gravitational wave  radiation and kilo-nova simulations associated with these mergers, predicting the Galactic Eu evolution is straightforward. Comparison with the observed run answers the title-question.

Since single star and binary population synthesis plays a crucial role a few facts are worth repeating.

\section{The massive star interacting binary fraction}
In the present paper the term `interacting close binary' stands for a binary where during the evolution of the components Roche lobe overflow/mass transfer will happen. For circular systems this means binaries with an orbital period less than (roughly) 10 years but it may be much larger when the orbit is eccentric. We further use the abbreviation MCB (massive close binary) and we distinguish O-type stars (ZAMS mass larger than $\sim$20 M$_\odot$) and B0-B3- (luminosity class V, IV and III)-type stars (ZAMS mass between 10 M$_\odot$ and 20 M$_\odot$). The mass ratio q = mass secondary/mass primary; the secondary/primary is the binary component with the smaller/larger mass on the ZAMS.

\citet{Garmany1980} studied the MCB fraction among all known O type stars brighter than m$_\mathrm{v}$ = 7 and north of -50$^\circ$ (a total of 67 O type single stars or primaries of binaries). They concluded that 33\% of the O type stars are the primary of a massive close binary (+/- 13\% accounting for small number statistics) with mass ratio q larger than 0.2 and period P < 100 days. When the period distribution is flat in the Log up to 10 years \citep{Popova1982, Mason1998} the real interacting O type binary fraction may be considerably larger.

The MCB fraction in the B0-B3 (luminosity class V, IV and III) spectral range has been investigated by \citet{Vanbeveren1998b,Vanbeveren1998c}. It was concluded that $\sim$32\% is primary of an interacting close binary. Notice the similarity with the MCB fraction of O-type stars. As for O-type stars also the B0-B3 MCB fraction can be considerably larger.

The study of the MCB fraction becomes much more complicated if one realizes that a significant number of the single stars in a massive star population may have had a binary history (e.g. single stars that became single after the SN explosion of the companion, or became single after the merger of both binary components) and the following procedure may therefore be mandatory.

Starting from the MCB fraction at birth (on the ZAMS) one calculates the content of the massive star population using a detailed massive star/binary population code. The question `what must be the massive close binary fraction at birth in order to explain the observed MCB fraction in a population where star formation happened continuously in time' was investigated by \citet{Vanbeveren1998a,Vanbeveren1998b,Vanbeveren1998c} and it was concluded that to explain the O and B0-B3 results discussed above, the MCB fraction at birth must be at least 70\%. Note however that in 1998 the physics of mergers was poorly understood. The MCB merger rate may therefore be larger than the 1998-value and as will be explained later on a model where the MCB fraction at birth equals 100\% cannot be excluded. Interestingly, about 14 years later the MCB fraction has been studied over again by \citet{Sana2012} and \citet{DeMink2013} who essentially confirmed the percentages discussed above. An interesting candidate of a single star with a binary history is $\zeta$ Pup. In the early eighties this star was considered as a gift from the gods as far as single star evolution is considered. However, the star is a runaway, e.g. it became runaway due to the supernova explosion of a companion or it became runaway due to the dynamical interaction in a dense cluster but also here at least one binary was involved in the formation. So $\zeta$ Pup may be a gift from the gods as far as binary evolution is concerned \citep[see also][]{Vanbeveren2012}.

The WR+OB binary fraction $\sim$30-40\% \citep{Vanbeveren1980} and many subsequent studies have confirmed this number. The O-type binaries are the WR+OB progenitors and we can try to answer the question `what must be the WR+OB progenitor binary fraction in order to explain the 30-40\% WR+OB binary fraction'. This was studied by \citet{Vanbeveren1998a,Vanbeveren1998b,Vanbeveren1998c} as well and it was concluded that a MCB fraction at birth >70\% also explains the observed WR+OB binary fraction.

Last not least, 15\% of the red supergiants seems to have a binary component \citep{Neugent2020} and also here this number is compatible with a >70\% MCB binary fraction at birth.

The population of early B-type supergiants deserves special attention. Figure 1 illustrates the HR-diagram of about 2500 massive stars in stellar aggregates in the Solar Neighborhood \citep[data taken from][]{Humphreys1984b}. When compared to massive single star evolution we conclude that the supergiant stars in the red box are post core hydrogen burning, hydrogen shell burning stars. These stars are expected to cross the box towards the red on the thermal Kelvin-Helmholts timescale which is very short (a factor 100-1000 shorter than the corresponding core hydrogen burning timescale). Therefore, compared to the number of stars in the core hydrogen burning box, the number of stars in the red box of Figure 1 should be 100-1000 smaller than in the core hydrogen burning box and by inspecting Figure 1 it is clear that this is far from being the case. Humphreys and McElroy also presented the results for the Magellanic Clouds and the effect discussed above is also visible there. Furthermore, only few stars in the red box seem to be close binaries (probably less than 20\%) which is odd at first glance considering the fact that the MCB fraction at birth of the progenitors of these supergiants is >70\%.

A famous example of the stars in the red box is the blue supergiant B-type progenitor of SN1987A, a star in the Large Magellanic Cloud with a mass $\sim$20 M$_\odot$. Binary models for SN1987A have been presented by \citet{Podsiadlowski1990,Podsiadlowski1992} and by \citet{DeLoore1992} following original suggestions of \citet{Hellings1983,Hellings1984}. These studies learned us that a post-core hydrogen burning star with a mass $\sim$20 M$_\odot$ with an under-massive helium core and an over-massive hydrogen shell will remain in the blue part of the HR-diagram during most of its core helium burning phase, populating in this way the red box of Figure 1. So, question: how to make a (massive) post-core hydrogen burning star with an under-massive helium core and an over-massive hydrogen shell? \citet{Vanbeveren2013} and \citet{Justham2014} proposed the merger of a massive case B close binary before the onset of core helium burning of the primary (case Br). In Figure 2 we show the evolutionary track of 3 binary models (we refer to our 2013 paper for a discussion of the evolutionary results and details about the computational method) where soon after the onset of RLOF both stars enter a contact phase from whereon the secondary merges with the primary. Merging is treated as a fast accretion process. At the end of the merging process the merger rapidly restores thermal equilibrium and becomes a core helium burning star but with a helium core that is under-massive with respect to its total mass. The merger remains a blue star till the end of core helium burning, e.g. these mergers populate the red box of Figure 1. Note that the merger is N overabundant, He overabundant, CO underabundant. Interestingly, \citet{Menon2024} obtained chemistry data for 59 blue supergiants in the Large Magellanic Cloud and compared them with the theoretically predicted chemistry of mergers. The authors conclude that for at least half of these supergiants the observations correspond to the merger predictions, a result that strongly support the merger model for these stars. Note that the merger model naturally explains why the blue supergiant binary fraction is so low.

The situation now is as follows: a significant percentage of case Br binaries merge and become single blue supergiants. A significant percentage of the MCBs evolve via a quasi-conservative RLOF and become WR+OB binaries. What makes the difference? And how important is the blue supergiant merger model when we repeat the exercise where we link the MCB fraction and the MCB fraction at birth as discussed above? We are prepared to stick out our neck: accounting for the observed large number of blue supergiants and the highly plausible merger model for these supergiants (which makes them single stars), repeating the exercise discussed above let us conclude that the MCB fraction at birth is close to 100\%. The simulations discussed in section 4 have been calculated with a MCB fraction at birth = 70\%. With 100\% the results change of course but the overall conclusions remain the same.

\begin{figure}
\centering
\includegraphics[width=14cm]{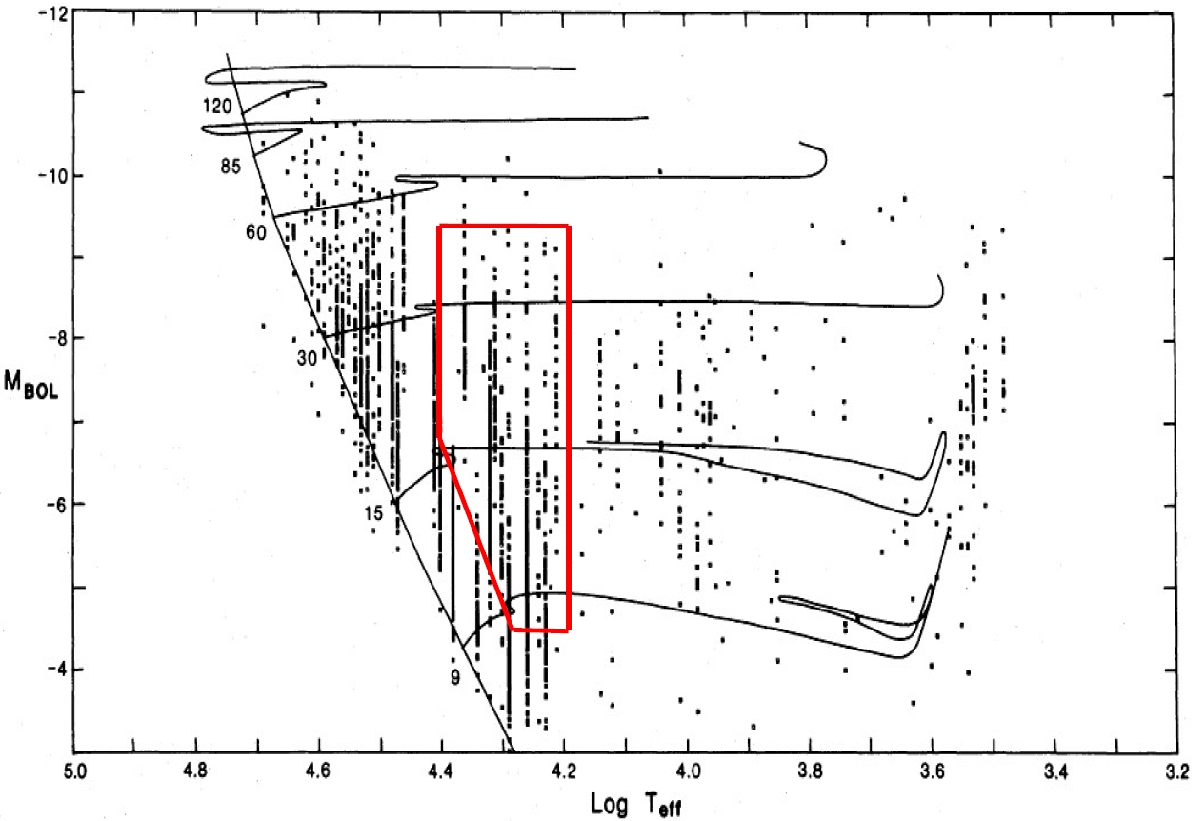}
\bigskip
\begin{minipage}{14cm}
\caption{The HR-diagram of 2500 massive stars in stellar aggregates in the Solar Neighborhood taken from \citet{Humphreys1984b}. The single star evolutionary tracks are Geneva tracks \citep{Schaller1992}. The red box contains the early B type supergiants discussed in the text.}
\end{minipage}
\end{figure}

\begin{figure}
\centering
\includegraphics[width=14cm]{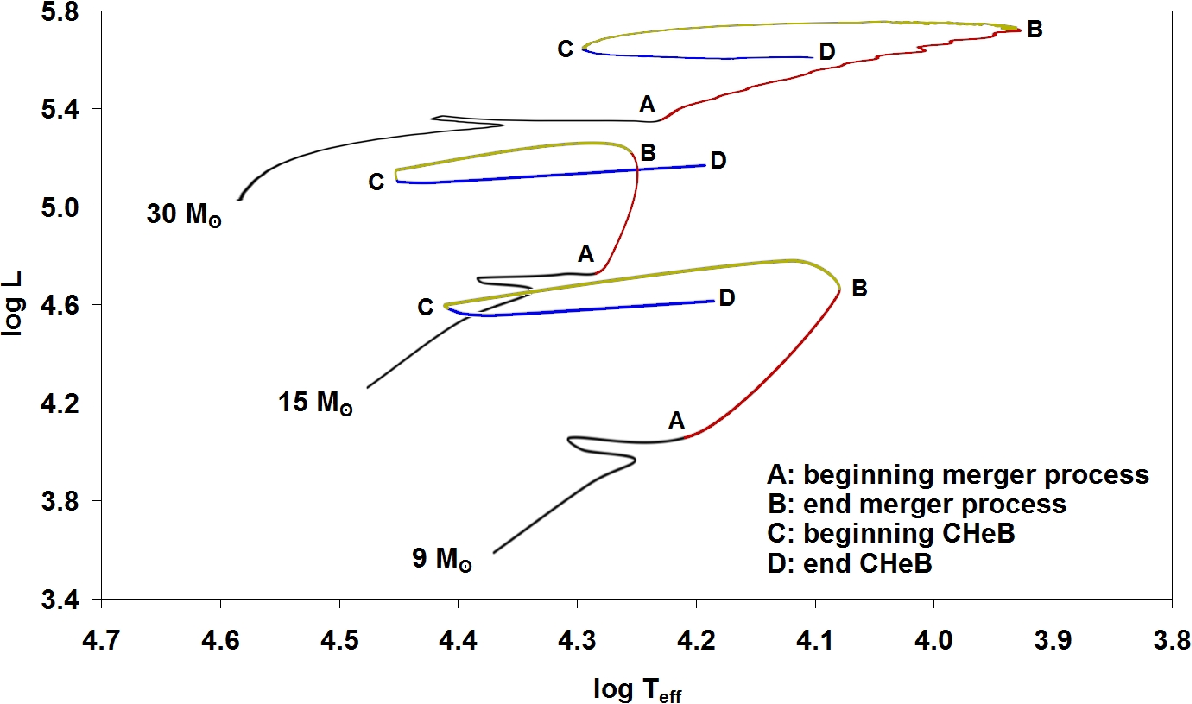}
\bigskip
\begin{minipage}{14cm}
\caption{Evolutionary tracks of 3 case B binary models (30+20M$_\odot$, 15+9M$_\odot$, 9+6.7M$_\odot$) where merging of the two stars starts at point A (merging is treated as a fast accretion process, where the secondary star is accreted onto the primary).}
\end{minipage}
\end{figure}

\section{The effect of LBV mass loss on the evolution of massive single stars and binaries}
Luminous Blue Variables (LBVs) have a luminosity Log L/L$_\odot$ > 5.4 and they occupy a region in the HR-diagram that corresponds to the hot part of the hydrogen shell burning phase of stars with an initial mass higher than 20 M$_\odot$. The observed LBV-mass loss rates range between 10$^{-5}$ and 10$^{-3}$ M$_\odot$/yr and on top of that LBVs are known to experience major mass eruption phases by a process that is as yet unclear. The observations reveal that the region in the HR-diagram which is located to the right of the region occupied by the LBVs with a luminosity Log L/L$_\odot$ > 5.5 (corresponding to stars with an initial mass higher than 30-40 M$_\odot$) is almost void. There are a few yellow hypergiants but no red supergiants. This is also observed in the Magellanic Clouds. Based on this fact and on the observed mass loss rates it was suggested by \citet{Humphreys1979,Humphreys1984a} that the LBV-mass loss has to be so high that it prohibits a hydrogen shell burning star with initial mass >30-40 M$_\odot$ from evolving into a red supergiant, and that this scenario is metallicity independent. From evolutionary point of view this is a very robust criterion allowing to calculate the LBV-mass loss rate that is needed. This was investigated by the Geneva stellar evolution team (\citet[]{Schaller1992}, see also \citet{Vanbeveren1998b, Vanbeveren1998c}) and it was concluded that when a massive star becomes an LBV and loses mass at a few times 10$^{-3}$ M$_\odot$/yr up to 10$^{-2}$ M$_\odot$/yr stars with an initial mass >30-40 M$_\odot$ lose their hydrogen rich layers and they become Wolf-Rayet stars without becoming RSGs first.

As discussed above it is conceivable that LBV stellar wind mass loss suppresses the redward evolution during hydrogen shell burning of stars with an initial mass higher than 30-40 M$_\odot$ and that stars in this mass range lose most of their hydrogen rich layers by this LBV mass loss process. As a consequence it cannot be excluded that the LBV mass loss rate suppresses the RLOF/common envelope phase in case Br/case Bc/case C binaries when the mass loser has a mass higher than 30-40 M$_\odot$ \citep[the LBV scenario of massive binaries as it was introduced by][]{Vanbeveren1991}. In these binaries the orbital period variation satisfies 

\begin{equation}
\frac{P}{P_0}=\left(\frac{M_{1,0}+M_{2,0}}{M_1+M_2}\right)^2
\end{equation}

where the subscript `0' stands for values at the beginning of the LBV mass loss phase.

Eq. (1) predicts a significant binary period increase during the LBV mass loss phase and we demonstrated in \citet{Mennekens2014} that this LBV scenario affects critically the predicted population of double-BH-binary mergers.

The results of the present paper (Table 1) with the label ``LBV 40'' (resp. ``LBV 20'') are calculated assuming an average LBV wind (the average of the eruption + intereruption mass loss) of a few 10$^{-3}$ M$_\odot$/yr up to 10$^{-2}$ M$_\odot$/yr that prohibits the redward evolution of stars with initial mass > 40 M$_\odot$ (resp. 20 M$_\odot$). Notice that the results presented here would be entirely similar when other average values would be adopted provided that these values are high enough so that they prevent the occurrence of the RLOF. The label ``LBV off'' correspond to results computed by switching off the LBV eruption mass loss so that the redward evolution for stars with initial mass > 20 M$_\odot$ is not suppressed and, when such a star has a close companion, RLOF/common envelope evolution  will happen.

As a note added in proof, \citet{Cheng2024} proposed a parameter model for eruptive mass loss of LBVs and implemented this in the MESA stellar evolutionary code. Their results essentially confirm the Geneva and Brussels simulations discussed above.

\section{Compact object mergers and r-process elements}
In \citet{Mennekens2014,Mennekens2016} we used a combination of the Brussels population synthesis code and Galactic chemical evolution model \citep[for an extended review see][]{DeDonder2004} to study the effect of binary star evolution on the production of r-process elements. These elements can be produced and released during the merger of two NSs or one NS and one BH, the latter being termed a `mixed' merger. The research question is whether these compact object mergers alone can produce enough r-process elements to match observations. The amount of r-process elements released due to a particular merger depends on the masses of both compact objects, and were previously based on the calculations by \citet{Korobkin2012}. These NSs and BHs are obviously the end products of (binary) star evolution, and their masses and distance (determining their merger timescale due to the emission of gravitational wave radiation) are determined by their previous evolution. In particular, we investigated the influence of still uncertain processes during the life of a binary star on the amount of r-process elements produced and the time at which they are released. The most important of these processes are the mass transfer efficiency during stable Roche-lobe overflow (RLOF) and the energy conversion efficiency during unstable common envelope (CE) evolution. As is common in binary evolution studies, we use the parameters $\beta$ and $\alpha$ to denote these properties. For RLOF, $\beta$ is the fraction of matter lost by the donor star that is actually accepted by the gainer star. $\beta=1$ thus denotes conservative RLOF, while $\beta<1$ implies non-conservative mass transfer. For CE evolution, $\alpha$ is defined as the efficiency of the transfer of orbital energy into escape energy of the CE. As can be seen from the results in \citet{Mennekens2014,Mennekens2016} these parameters (among others) have an enormous influence on the amount of r-process elements produced and the time at which they are released. Hence, these papers include a number of models which combine different assumptions and parameters. Models which produce results that are incompatible with observations can then be excluded. A number of observations can be used. First of all, the number of predicted Galactic double NS mergers must be high enough to match observations. For this, we use the lower limit of 3 mergers per Myr found by \citet{Kim2010}. Any models producing less mergers are eliminated. Second, models which produce too much r-process elements to be compatible with observations can be eliminated. Those which produce too few r-process elements (all of them, as will be seen later) can obviously not be eliminated, as there may be other sources of r-process element enrichment. Third, one can also look at the aLIGO rates for NS+NS and NS+BH mergers predicted by the models. This will be discussed later, as no observational aLIGO rates were available in 2014. The conclusion in 2014 was that some models could be eliminated on the grounds above, but none of the models succeeded in producing sufficient r-process elements through NS+NS and NS+BH mergers alone. A possible solution is of course to include r-process element production though other means, such as single star type II supernovae at low metallicity, which then gave a number of reasonable models. It is notable that for most or all of these models, the mixed mergers (NS+BH) provided a notable contribution to the r-process element enrichment of the same order of magnitude as the contribution of double NS mergers. All these conclusions are graphically illustrated in Fig 7 of \citet{Mennekens2014}.

Fig 2(b) of \citet{Kobayashi2023} presented a comparative study of various attempts at the theoretical reproduction of the observed [Eu/Fe] (representative for r-process elements) vs. [Fe/H] rates. The conclusion is that none of these models produce enough r-process elements to match observations. This comparison is based on the yields of a single NS+NS merger calculation, and assumes that the same yields are produced in NS+BH mergers (whereas, at least in 2014 with the Korobkin yields, we found those to be notably larger). It should be noted that the Brussels model used in this comparative study (named model 2 in our 2014 paper) is not our model with the highest allowable r-process element production. That is to say, there were models in our 2014 study with higher, but \citep[assuming the yields used by][]{Kobayashi2023} not too high, r-process element production. Also, the star formation rate assumed by \citet{Kobayashi2023} is quite different from our own, which means that their Fig 2(b) is more ``smeared out'' in the horizontal (temporal) direction than the panels in Fig 7 of \citet{Mennekens2014}.

Recently, the Brussels code was updated with the results from \citet{Just2015} to determine the amount of r-process elements released during a particular merger. These yields are based on a larger and more diverse number of merger calculations. A very large difference with the yields used before, is that with the Just yields, mixed mergers (NS+BH) hardly contribute to the r-process element enrichment, especially at higher metallicity. Hence, the [r/Fe] vs. [Fe/H] now almost only depends on the amount of double NS mergers. With the updated code, we have calculated a number of models consisting of various combinations of parameters. Table 1 summarizes these models. Some of these (the first five lines in Table 1) are the same models as in 2014, while others consist of new variations. The left hand side of the table denotes the input parameters of these various models.

With the same considerations as in 2014, models that produce insufficient Galactic NS+NS mergers and models that (at any time during the evolution) produce too much r-process elements can be rejected. Notable models that can thus be eliminated are those with $\beta=0.5$, as they produce too few Galactic mergers. Stable mass transfer thus has to be (near) conservative in order to match observations. Among the models with $\beta=1$, those with $\alpha=1$ (i.e. high common envelope efficiency) produce too much r-process elements at certain times and can thus also be eliminated. While models with lower CE efficiency ($\alpha=0.5$) can not strictly be excluded (as there may be other r-process element sources), the best match with observations is found for a moderate CE efficiency of $\alpha=2/3$. The [r/Fe] vs. [Fe/H] evolution predicted by this model is shown in Fig 3 of the present paper. The solid line represents the enrichment when including only NS+NS mergers. The dashed line includes the enrichment by both NS+NS and NS+BH mergers. This figure can thus be directly compared to the black lines in the panels of Fig 7 in \citet{Mennekens2014} which also provide this distinction. The dots are observed values as discussed in the same paper. In this case, there is an excellent match between prediction and observation at higher metallicity (i.e. the right half of the figure). For low metallicity (corresponding to the first $\sim$100 Myr of Galactic evolution) however, there is still insufficient r-process element production, and another source is needed to match observations. As in 2014, a suggestion (although certainly not the only one) could be the contribution from single star type II supernovae at low metallicity. It is obvious that due to the Just yields, the difference between the solid and dashed line in the figure is much smaller than between the corresponding lines in any of the panels in Fig 7 of \citet{Mennekens2014}. This demonstrates that with the Just yields, NS+BH mergers barely contribute to the r-process element enrichment, and only at very low metallicity. We now revisit the comparison by \citet{Kobayashi2023}. With our updated code using the Just yields, our [r/Fe] vs. [Fe/H] for $\beta=1$ and $\alpha=0.5$ (model 2 from 2014, second line in Table 1) now indeed looks very much like the one shown in Fig 2(b) of \citet{Kobayashi2023}. However others, such as the one shown in Fig 3 of the present paper, would lie much higher.

Thanks to the availability of aLIGO data, a third comparison can now be made which was not possible in 2014. Combined data from the first three aLIGO runs \citep{Abbott2023} seems to suggest that there is a near parity between the number of observed NS+NS and NS+BH mergers. To be exact, the Collaboration's analysis suggests that aLIGO has detected two NS+NS mergers and four NS+BH mergers (as well as 69 BH+BH mergers). Given the small numbers, one must obviously allow for a large statistical error. This ratio is obviously also a quantity that is predicted by our various theoretical models. The `best' model mentioned above, with $\beta=1$ and $\alpha=2/3$, which (apart from early times) satisfactorily reproduces the [r/Fe] vs. [Fe/H] rates, predicts 24 times more NS+BH merger detections than NS+NS merger detections. Regardless of small number statistics concerning the observations, this number appears to be too high. There are a number of possible solutions for this mismatch. First, one could assume that the limit for direct BH formation may be higher than our standard value of 40 M$_\odot$. A number of models has been calculated where this limit has been brought to 70 M$_\odot$ \citep{Schneider2021}. This way, the number of NS+BH mergers is greatly reduced, without much influencing the (already satisfactory) [r/Fe] vs. [Fe/H] evolution as the mixed mergers hardly contribute to the r-process element enrichment. A second possible solution (on its own or in combination with the first) is to assume that the supernova kick imparted on a BH may be larger than in our standard model. We normally use an expression \citep[described in][]{Mennekens2014} to calculate the fallback and resulting kick velocity during compact object formation, which results in almost 100\% fallback (and thus a very small kick) in the case of a BH. Now we have calculated a number of models where we assume zero fallback (and thus a large kick) in the case of BH formation. Lastly, one may also adapt the luminous blue variable (LBV) scenario. In our standard models, it is assumed that stars with an initial mass above 40 M$_\odot$ evolve according to this scenario. We now also present some calculations where this limit is lowered to 20 M$_\odot$, as well as some others where this LBV scenario is avoided altogether even by massive stars \citep[denoted ``off'' in Table 1, a possibility also discussed in][]{Mennekens2014}. As can be seen in the right half of the table, most of these models result in a [r/Fe] vs. [Fe/H] evolution that is acceptable (and thus all very similar to the one shown in Fig 3). That is to say, not too high at any time, and not too low at high metallicity. The [r/Fe] lying too low at low metallicity is present in each model, and thus as mentioned before requires an alternative r-process element source. Table 1 also shows the ratio of expected NS+BH/NS+NS aLIGO detections, which as a reminder is expected to be about two, but may statistically lie anywhere between 0.33 and 15. Also the predicted amount of Galactic NS+NS mergers (with its observational lower limit of three per Myr) is shown. It should also be noted (as shown in Table 1) that most of our models predict no aLIGO BH+BH merger detections from a binary evolutionary origin. The only exception are the models with LBV off, but most of those then immediately have a BH+BH merger detection rate that is much higher (compared to NS+NS and NS+BH merger detections) than observations suggest. Therefore we certainly keep an open mind concerning other pathways to BH+BH mergers that have been suggested in the literature over the past decade.

The conclusion remains that none of our models is able to reproduce the observed amount of r-process element enrichment at early stages in Galactic evolution (at low metallicity) through compact object mergers alone, and thus certainly not when including only double NS mergers. Other sources of r-process elements, such as single stars, must be included to obtain an acceptable result. The [r/Fe] vs. [Fe/H] at later times (higher metallicity) can be reproduced more satisfactorily with the Just yields than with the Korobkin yields used by us in 2014. However, they remain extremely sensitive to binary star evolutionary parameters such as $\beta$, $\alpha$ and assumptions about BH formation and LBV evolution.

\begin{figure}
\centering
\includegraphics[width=14cm]{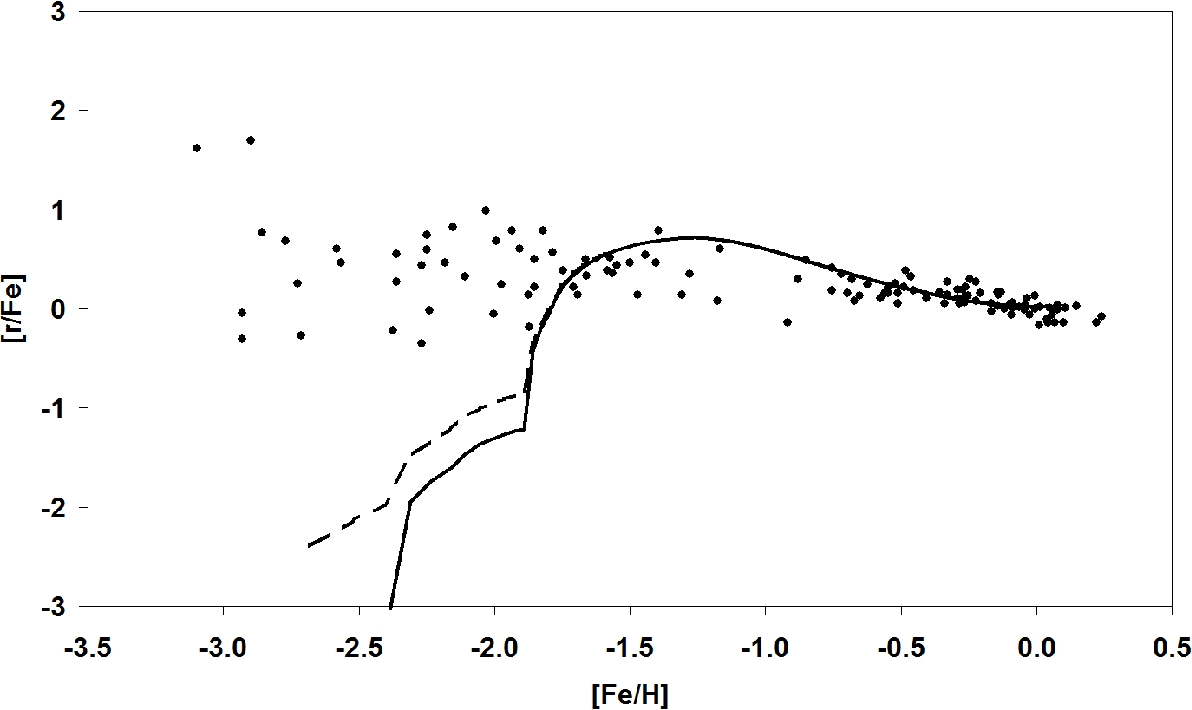}
\bigskip
\begin{minipage}{14cm}
\caption{r-process element evolution for a model with $\beta=1$, $\alpha=2/3$ and other assumptions standard (= sixth line in Table 1). Solid line is enrichment due to NS+NS only, dashed line due to NS+NS and NS+BH combined. Dots are observations.}
\end{minipage}
\end{figure}

\begin{table}
\centering
\begin{minipage}{12cm}
\caption{Overview of models computed in this study. Left part contains input parameters, right part contains output predictions. `std' denotes the standard fallback mechanism (see text).}
\end{minipage}
\bigskip
\begin{tabular}{ccccc|cccc}
\hline
 & & BH & BH & LBV & [r/Fe] & aLIGO & aLIGO & Gal NS+ \\
 & & lim & fall- & lim & vs. & $\frac{\mathrm{NS+BH}}{\mathrm{NS+NS}}$ & $\frac{\mathrm{BH+BH}}{\mathrm{NS+NS}}$ & NS merg \\
$\beta$ & $\alpha$ & (M$_\odot$) & back & (M$_\odot$) & [Fe/H] & ratio & ratio & (/Myr) \\
\hline
1 & 1 & 40 & std & 40 & high & 17.7 & 0 & 42.1\\
1 & 0.5 & 40 & std & 40 & low & 29.1 & 0 & 8.65\\
0.5 & 1 & 40 & std & 40 & low & 350 & 0 & 2.05\\
1 & 1 & 40 & std & off & high & 9.48 & 315 & 42.1\\
1 & 0.5 & 40 & std & off & low & 67.5 & 1880 & 8.65\\
1 & 2/3 & 40 & std & 40 & ok & 24.3 & 0 & 19.0\\
1 & 2/3 & 70 & std & 40 & ok & 11.2 & 0 & 19.0\\
1 & 2/3 & 40 & 0 & 40 & ok & 11.1 & 0 & 19.0\\
1 & 2/3 & 70 & 0 & 40 & ok & 3.73 & 0 & 19.0\\
1 & 2/3 & 40 & std & off & ok & 23.3 & 820 & 19.0\\
1 & 2/3 & 70 & std & off & ok & 10.5 & 27.3 & 19.0\\
1 & 2/3 & 40 & 0 & off & ok & 9.72 & 158 & 19.0\\
1 & 2/3 & 70 & 0 & off & ok & 3.64 & 3.71 & 19.0\\
1 & 2/3 & 40 & std & 20 & ok & 8.12 & 0 & 19.4\\
1 & 2/3 & 70 & std & 20 & ok & 3.41 & 0 & 19.5\\
1 & 2/3 & 40 & 0 & 20 & ok & 3.42 & 0 & 19.4\\
1 & 2/3 & 70 & 0 & 20 & ok & 0.96 & 0 & 19.5\\
\hline
\end{tabular}
\end{table}

\bibliographystyle{bullsrsl-en}
\bibliography{D_Vanbeveren}

\end{document}